# ARTICLE INFORMATION

**Article title**

A Dataset of Low-Rated Applications from the Amazon Appstore for User Feedback Analysis

**Authors**

Nek Dil Khan[1], Javed Ali Khan[2], Darvesh Khan[3], Jianqiang Li[1]*, Mumrez Khan[4], Shah Fahad Khan[5]

**Affiliations**

[1]Faculty of Information Technology, Beijing University of Technology, Beijing, 100124, China; nekdilkhan@emails.bjut.edu.cn, lijianqiang@bjut.edu.cn

[2]Department of Computer Science, School of Physics, Engineering, and Computer Science, University of Hertfordshire, Hatfield, AL10 9AB, UK; j.a.khan@herts.ac.uk

[3]Department of Software Engineering, CECOS University of IT & Emerging Sciences, Peshawar, Khyber Pakhtunkhwa, Pakistan darwesh.khan.bsse-2021b@cecosian.edu.pk

[4]Faculty of Computer Science and Engineering, Xi'an University of Technology, Xian, 710121, China; mumrez.khan@stu.xaut.edu.cn

[5]Computer Software Engineering Collage, Dalian University of Technology, Dalian 116000, China
sfahad@mail.dlut.edu.cn

**Corresponding author's email address and Twitter handle**

Jianqiang Li: lijianqiang@bjut.edu.cn ; Javed Ali Khan: j.a.khan@herts.ac.uk

**Abstract**

In today's digital landscape, end-user feedback plays a crucial role in the evolution of software applications, particularly in addressing issues that hinder user experience. While much research has focused on high-rated applications, low-rated applications often remain unexplored, despite their potential to reveal valuable insights. This study introduces a novel dataset curated from 64 low-rated applications sourced from the Amazon Software Appstore (ASA), containing 79,821 user reviews. The dataset is designed to capture the most frequent issues identified by users, which are critical for improving software quality. To further enhance the dataset utility, a subset of 6000 reviews was manually annotated to classify them into six district issue categories: user interface (UI) and user experience (UX), functionality and features, compatibility and device specificity, performance and stability, customer support and responsiveness, and security and privacy issues. This annotated dataset is a valuable resource for developing machine learning-based approaches aiming to automate the classification of user feedback into various issue types. Making both the annotated and raw datasets publicly available provides researchers and developers with a crucial tool to understand common issues in low-rated apps and inform software improvements. The comprehensive analysis and availability of this dataset lay the groundwork for data-derived solutions to improve software quality based on user feedback. Additionally, the dataset can provide opportunities for software vendors and researchers to explore various software evolution-related activities, including frequently missing features, sarcasm, and associated emotions, which will help better understand the reasons for comparatively low app ratings.

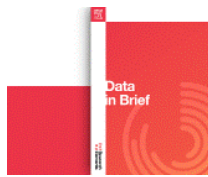

## Data SPECIFICATIONS

Table 1 lists the data specifications for enhanced clarity.

**Table 1:** Specifications

| Specification | Details |
|---|---|
| **Subject** | Computer Science, Software Engineering |
| **Specific Subject Area** | Software Evolution, Crowd-based RE, Natural Language Processing (NLP), User experience (UX) |
| **Type of Data** | Tabular |
| **Data Collection** | Automated web scraping using the Google Chrome extension Instant Data Scraper that extracts data from web pages and exports it as Excel or CSV files. Instant Data Scraper: |
| **Data Source Location** | Beijing University of Technology. No.100 Ping Le Yuan, Chaoyang District, Beijing City, PR China, 100124. |
| **Data Accessibility** | The raw and annotated datasets are published, whose details are provided below. Repository name: Mendeley Data identification number: 10.17632/zztwrn36n8.2 Direct URL to the data: https://data.mendeley.com/datasets/zztwrn36n8/1 GitHub Link: https://github.com/nekdil566/Dataset-of-User-Reviews-from-Low-Rated-Amazon-Appstore-Applications/blob/main/README.md |
| **Data Volume** | Raw dataset (12.1 MB), annotated dataset () |
| **Data Quality Assurance** | The dataset undergoes a rigorous assurance process to ensure both consistency and accuracy. Additionally, it is cleaned to remove duplications, inconsistencies, |
| | or irrelevant information, ensuring it remains comprehensive and precise. The annotated reviews are verified by multiple coders to achieve high inter-coder agreement, further enhancing the dataset reliability for ML applications. The dataset provides high-quality, manually labelled data suitable for advanced analytical tasks such as issue detection, opinion, and sentiment analysis. |
| **Potential Uses** | The dataset can enhance software quality by pinpointing common user issues, which facilitates targeted updates and feature enhancements. They are valuable for training Machine Learning (ML) models to classify user feedback and automatically identify recurring issues. Furthermore, the dataset supports sentiment analysis, UX research, and market analysis, providing insights into user dissatisfaction and informing more user-centered software development. |
| **Related Research Article** | Mining software insights: uncovering the frequently occurring issues in low-rating software applications [1]. |

## VALUE OF THE DATA

- **Dataset Use Case:** The dataset comprises a comprehensive collection of end-user reviews, specifically 79821 reviews collected from the Amazon Play Store for 64 apps across 14 categories. Unlike previous approaches, the dataset focuses on comparatively low-rated applications, enabling software vendors, developers, and researchers to concentrate on frequently occurring problems that contribute to low ratings. For this purpose, the dataset was utilized to identify frequently occurring issues in app stores by analyzing end-user feedback using fine-tuned machine learning and deep learning algorithms [1]. Also, the customized dataset is used to identify end-user emotions embedded in the reviews by employing ChatGPT and fine-tuned DL algorithms [3]. By making this dataset publicly accessible, it empowers the research community to explore the root causes of negative feedback, which can inform more effective software improvements and ultimately enhance the overall user experience. Additionally, the customized dataset was used first to identify frequently occurring issues, as determined by part-of-speech analytics. Then, associated emotions were identified using fine-tuned ML classifiers, thereby enhancing the understanding of end-user grievances [2]. Moreover, it is reported that a single review can encompass multiple issues, such as performance and security. Therefore, it is essential to re-annotate the dataset for a multi-label classification problem, which to date hasn't been explored in software engineering literature, according to our knowledge. These models can streamline the process of detecting useful information in end-user feedback, providing developers with a scalable tool for quickly identifying areas that require attention.
- **Enabling Root Cause Analysis of User Dissatisfaction:** The research dataset enables researchers and software vendors to employ topic modeling techniques such as Latent Dirichlet Allocation (LDA) to uncover latent themes in negative feedback, allowing them to identify systematic issues behind low app ratings. It will also help to validate the research findings [1] by identifying frequent issues using the LDA algorithm. This analysis will help in eliminating the data annotation process, which is time-consuming and challenging. Additionally, researchers can link thematic clusters to functional components, supporting empirical studies on software quality and user-perceived defects, and propose solutions to improve the quality of existing low-rating software apps.

Moreover, the dataset can be reannotated at the sentence level and compared with its performance at the review level for issue identification.

- **Facilitating ML Model Development and Validation:** The dataset comprises 79821 user reviews, amongst which 6000 are manually annotated into six frequently occurring issue types in low-rating software apps. The dataset can serve as a standard for training and evaluating supervised and semi-supervised classification models. The dataset can be reannotated by further understanding the possible low ratings of software apps. For example, an automated pipeline can be developed to identify sarcasm and its types (positive and negative) to improve the performance of existing sentiment analysis approaches [4]. Similarly, the low-rating dataset can be explored for fine-grained new features, as we argue that users might have reported some missing features, which might have resulted in low ratings. Similarly, the dataset provides a reliable basis for validating and comparing different classification frameworks used in crowd-based requirements engineering. For this purpose, the labelled dataset can be utilized to identify the generalizability of the research approaches by conducting ablation studies. Researchers can reuse the data to fine-tune the Transformer or Large Language Models (LLM) for domain-specific app review analysis, improving model generalization in low-resource settings.

- **Issues or Features Priorities for Software Evolution:** In the literature, many approaches have been proposed, which classify reviews into frequently occurring issue types or features. However, some issues or features are more critical than others and need immediate response from the developers. For this purpose, the dataset can be used first to identify the frequently occurring issues or features mentioned by the end-users using a natural language processing pipeline, part-of-speech tagging, and topic modeling. Then, quantify how many end-user reviews mentioned each issue or feature by adding its co-occurrence. A tool can be provided to support software developers in ranking frequently mentioned issues or features for immediate action. Furthermore, sentiment analysis can be added to frequently identify issues or features, thereby enhancing the understanding of user context and opinion regarding the reported issues and features, ultimately improving ranking.

- **Informing Educational Curriculum in Software Engineering and UX:** Recently, Artificial Intelligence (AI) and ML have been added to the software engineering and computer science curriculum due to their widespread application and success. Similarly, Requirements gathering is not limited to in-house users; instead, for market-based applications, requirements need to be gathered automatically by analyzing end-user feedback from social media platforms. This dataset offers academics opportunities to utilize it in demonstrating various CrowdRE concepts, such as developing a basic pipeline for classifying reviews into various requirements-related information, including features, issues, and non-functional requirements, by re-annotating the dataset to better demonstrate the machine learning cycle. It fosters experiential learning in requirements engineering, usability engineering, and software evolution modules. Students can gain hands-on experience with authentic feedback, learning how poor UX, performance issues, or lack of support contribute to app failure [1].

- **Improved Customer Service:** In the literature, there has been evidence of increasing end-user feedback against the software apps in the app stores and other social media platforms. It becomes challenging and laborious to respond to each end-user feedback manually on these platforms for improved customer service. However, the dataset can serve as a ground truth in developing an automated and enhanced customer service system by analyzing end-user feedback using NLP to extract issue type. Then use a classification approach to classify it into predefined categories, such as bugs, performance, security, etc. After identifying the issue type, a predefined message can be fetched from a dictionary associated with this category. Moreover, advanced NLP can be applied to make the messages more personalized by extracting the user's name or acknowledging the identified issue. Additionally, state-of-the-art transformers, such as GPT, BERT, or T5, can be used to provide contextual understanding of end-user reviews by detecting the issue type, associated emotion, tone, and intent. Customized responses should be added automatically based on this information in an improved and timely manner, which will help retain customers and improve quality.

## BACKGROUND

In the competitive realm of digital marketplaces, such as the ASA, maintaining high-end user satisfaction poses a significant challenge for developers. Recent studies have consistently shown that analyzing user feedback is a vital alternative source of gaining valuable insights for enhancing software quality and functionality for existing in-house stakeholders. While all end-user feedback is useful, reviews of low-rated applications are particularly critical as they often contain detailed accounts of software failures, usability issues, and unmet user needs in addition to the missing features. By focusing specifically on these instances of user dissatisfaction, the research dataset offers a rich foundation for a detailed analysis of the factors that lead to negative application ratings. This study builds upon foundational research and employs NLP and ML techniques to extract actionable insights from user feedback [1] [2]. By providing a dataset of low-rated apps to existing datasets, it creates a crucial feedback loop for developers in understanding comparatively low ratings of software apps. It serves as a valuable resource for training and validating a new generation of analytical models that aim to advance user-centric software development.

## DATA DESCRIPTION

The dataset was organized into a comprehensive Table detailing 64 software applications across 14 categories in the Amazon App Store. Each entry in the dataset provides specific information about the application, including its category, name, and number of end-user reviews, as shown in Table 2. The dataset structure is as follows.

**Column 1 (Serial Number (S/no.)):** This column sequentially numbers entries ranging from 1 to 14, each representing a unique application category.

**Column 2 (App category):** This column lists the application categories. There are 14 categories in total, including Business Apps, Communication Apps, Education Apps, Food and Drinks Apps, Game Apps, Movies and TV Apps, Music and Audio Applications, Novelty Applications, Photos and Videos Apps, Productivity Apps, Sports and Exercise Applications, Utilities Apps, Lifestyle Apps, and Travel Apps. These categories were selected to make the dataset more generalized, representing the major app categories.

**Column 3 (Software Applications):** Specific applications are listed under each category. For example, applications like Hammer Print, Sketch Guru, Sketchbook, Logo Maker, and Logo Creator are listed under Business Apps. Each category contained various applications, totaling 64 applications across all categories.

**Column 4 (End-user Review):** This column provides the number of end-user reviews for each application category. These figures provide an overview of user engagement and feedback for each category.

Table 2. Summary of the Dataset [1].

| S/No | Apps category | Software Applications | End-user Review |
|---|---|---|---|
| 1 | Business Apps | 1: Hammer print, 2: Sketch guru, 3: Sketch book, 4: Logo maker & logo creator | 11034 |
| 2 | Communication Apps | 1: Just talk, 2: Texteme, 3: Skype 4: Add Me Snaps, 5: Free text | 5922 |
| 3 | Education Apps | 1: TED TV, 2: Casper's company, 3: World now, 4: Amazon silk: duolingo.com | 6968 |
| 4 | Food and Drinks Apps | 1: Food network 2: Food Planner, 3: Italian recipes, 4: Chef tap: recipe clipper, planner | 5521 |
| 5 | Game Apps | 1: Mobile strike, 2: Drive car spider simulator, 3: Chapters: interactive stories, 4: Crazy animal selfie lenses, 5: Cradle of empires-match | 4748 |
| 6 | Movies and TV Apps | 1: Anthem TV–It's free cable, 2: Hauppauge my TV, 3: Stream TV, 4: Trashy movies channel, 5: TV Player watch live TV & on-demand | 5982 |
| 7 | Music and Audio Applications | 1: Karaoke party by red karaoke, 2: Mp3 music download, 3: Red karaoke sing & record, 4: Voice changer, 5: Kara Fun karaoke & singing | 1889 |
| 8 | Novelty Applications | 1: Ghost radar: classic, 2: Xray scanner, 3: Cast manager, 4: Age scanner, 5: Clock in Motion | 3104 |
| 9 | Photos and videos Apps | 1: Air Beam TV screen mirroring receiver, 2: Air Screen, 3: Snappy photo filter and stickers, 4: Screen Cast, 5: Rec Me free screen recorder | 5089 |
| 10 | Productivity Apps | 1: PDF max pro–read, 2: Floor plan creator, 3: Office Suite free | 5421 |
| 11 | Sports and Exercise Applications | 1: FOX Sports: Stream live NASCAR, boxing, 2: fubo TV-watch live sports, TV shows, movies & news, 3: NBC Sports, 4: CBS sports stream & watch live | 10630 |
| 12 | Utilities Apps | 1: Print Bot, 2: Floor plan creator, 3: tinyCam monitor free, 4: Tv screen mirroring, 5: Optimizer & trash cleaner tool for kindle fire | 3689 |
| 13 | Lifestyle Apps | 1: Screen mirroring, 2: DOGTV, 3: Fanmio boxing, 4: 3d Home designs layouts, 5: Home design 3D-free | 4303 |
| 14 | Travel Apps | 1: World Webcams,2: Compass 3: Tizi World My Pretend Play Town for Kids 4: My Route Planner Travel 5: Public transport maps offline | 5521 |
| Total apps | | 64 | 79821 |

The dataset is presented in a tabular format for easy understanding and analyses. A few instances from the dataset are shown in Table 3 to illustrate its structure. The dataset was organized into four primary columns:

**Rating:** This column captures the quantitative star ratings assigned by the user (e.g.,"1.0 out of 5 stars"). It serves as the primary metric for gauging user satisfaction with each review.

**Title of Review:** This column contains the headings or summaries of the review provided by the user, which often encapsulate the main points of their feedback.

**Full Review:** This is the most detailed column, containing the complete text of user feedback. This is the primary source of qualitative and quantitative analysis.

**Issue Type:** This column identifies the type of issue raised by the user within their review. Each review is categorized based on predefined issue types, including Performance & Stability, UI & UX, Functionality & Features, Compatibility & Device-Specific Issues, Customer Support & Responsiveness, and Security & Privacy Issues. These issue types are identified based on their frequency in the end-user comments.

Table 3. Dataset Attributes

| Rating | Review Title | Base Review | Issue Type |
| --- | --- | --- | --- |
| 1.0 | In all fairness maybe if I had a wireless printer I would have had better | Could not get this function to work, took way too much effort to be unsuccessful, in all fairness maybe if I had a wireless printer I would have had better luck | Functionality and Features |
| 1.0 | Junk | Does not work at all - even after purchasing upgrade. Paper runs through printer with no print whatsoever! No number to call for refund or support. Got ripped off period. AVOID! | Customer support and responsiveness |
| 1.0 | Not for Android? | Unable to get it to print from any of my android devices. Difficult to sync my printer with the app, and until would not work. | Compatibility and device |
| 1.0 | Frequent Crashes and Freezing Make app unusable | App freezes every time I try to load a document. Even after updating, it still has issues with crashing and freezing, making it practically useless. | Performance and stability |
| 1.0 | Too many privacy concerns | Too many privacy concerns! The app asks for unnecessary permissions and even tracks my location without clear consent. I won't trust it with my data. | Security and privacy |
| 1.0 | Very Confusing | Extremely difficult to navigate, especially for beginners. Buttons are poorly labeled and I struggled to find basic functions like saving files or changing settings. | User interface and UX |

# EXPERIMENTAL DESIGN, MATERIALS AND METHODS

The data acquisition process for this dataset involved a systematic approach to collect and categorize information about low-rated software applications from the Amazon Appstore. The methodology is as follows:

**1. Selection Criteria for Applications**

**Rating Threshold:** A Specific rating threshold was established, selecting applications with user ratings of three stars (out of five) or lower. This ensured that the chosen applications were rated significantly lower than the average for their respective categories.

**Minimum Review Count:** Applications with numerous reviews were chosen to ensure that a low rating indicated a broader user opinion rather than a few isolated views.

**2. Data Collection Process:**

**Automated Scraping Tool:** A publicly available Google Chrome extension, Instant Data Scraper https://chromewebstore.google.com/detail/instant-data-scraper/ofaokhiedipichpaobibbnahnkdoiiah?hl=en-US&utm_source=ext_sidebar was used to automate the data extraction process from the Amazon App Store. This tool extracts data from web pages and exports them into structured file formats.

**Data points collected**: For each application, the scraping tool was configured to collect the rating, title of the review, and full review, along with other relevant metadata.

**3. Data Categorization:**

**Manual Review:** After the initial automated collection, each application was manually reviewed to ensure that it met the selection criteria. The applications were then organized into one of the 14 predefined categories based on the primary function.

**4. Data Verification and Cleaning:**

**Accuracy Check:** The collected data underwent a verification process to ensure accuracy, which involved cross-checking the scraped data with live information on the Amazon App Store.

**Data Cleaning:** The dataset was cleaned to remove duplications, inconsistencies, and irrelevant information, a crucial step in maintaining the quality and reliability of the data.

**5. Data Compilation:**

**Tabulation:** The final, cleaned data were compiled into a structured table format using Microsoft Excel for ease of use and accessibility.

**Formatting:** The data were organized with columns for the user rating (stars), title of the reviews, and full review text. It was prepared for final distribution in CSV format for broad compatibility.

# LIMITATIONS

This study offers valuable insights into user feedback from low-rated applications on the ASA store; however, several limitations should be considered. One significant constraint is the restricted access to end-user reviews due to recent changes in Amazon's review policies. Initially, the automated web scraping tool allowed comprehensive access to a wide array of reviews, but now the tool can only retrieve a maximum of 10 pages per app. This limitation affects the dataset's coverage, potentially omitting user concerns mentioned in reviews beyond the 10-page threshold. As a result, the data may not fully represent all issues, particularly those that are less frequently mentioned. This reduction in available data could lead to an incomplete understanding of the common user complaints across a broader range of reviews. The data was collected in April 2023, and further details on Amazon's review policies can be found on their official Developer Policy https://developer.amazon.com/.

Additionally, the dataset's generalizability is confined to the ASA store, representing only a small portion of the global app market. User behavior and review patterns can vary significantly across different platforms, making the findings from the dataset potentially less applicable to apps on other stores. The dataset also captures static snapshots in time, which means that the dynamic nature of app updates and changes may not be reflected in the collected data, potentially altering user ratings and feedback. Furthermore, the study's focus on low-rated apps with a minimum of 350 reviews excludes insights from newer or slightly higher-rated applications, which could still offer valuable information. The reliance on automated scraping tools introduces another challenge, as any changes to the app store's website layout could disrupt data collection and lead to errors. Lastly, the manual categorization of issues, although systematically performed, involves an element of subjective interpretation that may influence the final classifications, despite efforts to maintain objectivity.

# ETHICS STATEMENT

The dataset was compiled and processed with strict adherence to ethical considerations. The primary data source, the Amazon Appstore, is a public domain where application information is available to users. No private or sensitive user data was extracted or used for data collection. This study focused solely on publicly accessible data, specifically application names, categories, and the number of user reviews, which do not involve any personal or confidential user information.

Furthermore, this study did not involve human or animal subjects, and there was no interaction with individuals or groups. Therefore, ethical approval was not required for this study. Data collection and analysis were conducted to maintain the highest standards of integrity and respect for data sources. The dataset and methods used for its compilation are intended solely for academic and research purposes. They are designed to contribute to the understanding of user preferences and trends in the digital marketplace, particularly in the context of low-rated software applications. Any potential use of this dataset is expected to uphold these ethical standards, ensuring that the data are used responsibly and appropriately in subsequent research or analysis.

# CRediT AUTHOR STATEMENT

**Nek Dil Khan:** Methodology, Conceptualization, Writing – original draft, Writing – review & editing.
**Javed Ali Khan:** Methodology, Writing – original draft. **Javed Ali Khan:** Methodology, Writing – original draft  **Darvesh Khan:** Software, Data curation, Investigation, Writing – review & editing. **Mumrez Khan:** Software, Validation, Writing – review & editing. **Shahfahad Khan:** Software, Validation, Writing – review, and editing.

# DECLARATION OF COMPETING INTERESTS

The authors declare that they have no financial interests or personal relationships that could have influenced the outcome of this work.

# Data Availability

*Data identification number: 10.17632/zztwrn36n8.1*

Direct URL to the data: https://data.mendeley.com/datasets/zztwrn36n8/1

GitHub Link: https://github.com/nekdil566/Dataset-of-User-Reviews-from-Low-Rated-Amazon-Appstore-Applications/blob/main/README.md

## REFERENCES

[1] N. D. Khan, J. A. Khan, J. Li, T. Ullah, and Q. Zhao, "Mining software insights: uncovering the frequently occurring issues in low-rating software applications," *PeerJ Comput Sci*, vol. 10, p. e2115, Jul. 2024, doi: 10.7717/PEERJ-CS.2115.

[2] N. Khan, J. Khan, J. Li, T. Ullah, … A. A.-I., and undefined 2024, "How do crowd-users express their opinions against software applications in social media? A fine-grained classification approach," *ieeexplore.ieee.orgND Khan, JA Khan, J Li, T Ullah, A Alwadain, A Yasin, Q ZhaoIEEE Access, 2024•ieeexplore.ieee.org*, Accessed: Sep. 18, 2024. [Online]. Available: https://ieeexplore.ieee.org/abstract/document/10591990/

[3] N. Khan, J. Khan, J. Li, … T. U.-E. S. with, and undefined 2025, "Leveraging Large Language Model ChatGPT for enhanced understanding of end-user emotions in social media feedbacks," *bdbanalytics.irND Khan, JA Khan, J Li, T Ullah, Q ZhaoExpert Systems with Applications, 2025•bdbanalytics.ir*, Accessed: Jun. 26, 2025. [Online]. Available: https://bdbanalytics.ir/wp-content/uploads/2025/02/Leveraging-Large-Language-Model-ChatGPT-for-enhanced-understanding-of-end-user-emotions-in-social-media-feedbacks.pdf

[4] E. Fatima, H. Kanwal, J. A. Khan, and N. D. Khan, "An exploratory and automated study of sarcasm detection and classification in app stores using fine-tuned deep learning classifiers," *Automated Software Engineering*, vol. 31, no. 2, pp. 1–53, Nov. 2024, doi: 10.1007/S10515-024-00468-3/METRICS.